\documentclass[manuscript,nonacm]{acmart}

\usepackage{subcaption}
\usepackage{adjustbox}
\usepackage{todo}
\usepackage{xcolor,colortbl}
\usepackage{snapshot}

\AtBeginDocument{  \providecommand\BibTeX{{    \normalfont B\kern-0.5em{\scshape i\kern-0.25em b}\kern-0.8em\TeX}}}

\definecolor{Draft}{rgb}{0.8,0.1,0.1}
\definecolor{red}{rgb}{1.0,0.0,0.0}
\definecolor{blue}{rgb}{0.0,0.0,1.0}

\definecolor{g4}{gray}{0.90}
\definecolor{g3}{gray}{0.99}
\definecolor{g2}{gray}{0.90}
\definecolor{g1}{gray}{0.84}
\definecolor{g0}{gray}{0.70}

\begin{document}

\title[What Holds Back Brain-Computer Interfaces?]{What Holds Back Brain-Computer Interfaces? Uncovering Challenges and Opportunities in BCI-controlled Games for Cerebral Palsy Rehabilitation }

\author{Bastian~Ilsø~Hougaard}
\affiliation{  \institution{Aalborg~University}
  \streetaddress{}
  \city{Aalborg}
  \country{Denmark}}
\email{biho@create.aau.dk}

\author{Kirstine~Schultz~Dalgaard}
\affiliation{  \institution{Aalborg~University}
  \streetaddress{}
  \city{Aalborg}
  \country{Denmark}}

\author{Kirstine~Johanne~Stougaard~Klebæk}
\affiliation{  \institution{Aalborg~University}
  \streetaddress{}
  \city{Aalborg}
  \country{Denmark}}

\author{Hendrik~Knoche}
\affiliation{  \institution{Aalborg~University}
  \streetaddress{}
  \city{Aalborg}
  \country{Denmark}}
  
\author{Mads~Jochumsen}
\affiliation{  \institution{Aalborg~University}
  \streetaddress{}
  \city{Aalborg}
  \country{Denmark}}

\renewcommand{\shortauthors}{...}

\begin{abstract}
Brain-computer interfaces (BCIs) offer promising avenues for cerebral palsy (CP) rehabilitation at home and in the clinic, using games that promote engagement and sustained training effort.
Nonetheless, the design constraints of BCI-based CP rehabilitation remain unclear, especially how individuals with CP experience a sense of control through BCI, and how they experience computer‑mediated game assistance.
To address this gap, we present preliminary clinical and user perspectives on BCI-based CP rehabilitation, drawing on in-clinic insights from a CP therapist and experiential accounts from ten individuals with CP engaging with BCI game prototypes.
Sporadic help in BCI games eased monotony, but also fostered doubts regarding agency. The therapist saw BCI rehabilitation as complementary to traditional training, facilitating the transition from playful exercises to autonomous, self‑managed training. We outline key challenges and opportunities to inform and empower further design and research of BCI training for CP.

\end{abstract}


\setcopyright{none}

\maketitle

\section{Introduction}
Brain–computer interfaces (BCIs) have been proposed as a technically feasible way to facilitate motor training for people with cerebral palsy (CP)~\citep{jochumsen_comparison_2025}. BCI training involves many repetitions making it monotonous, but adding games can improve people's engagement and motivation to train~\citep{scherer_game-based_2015}. However, human-computer interaction (HCI) research has yet to systematically delineate the interaction constraints imposed by BCI‑mediated CP motor training, particularly how people with CP experience BCI controlled games, and whether computer-mediated assistance can compensate for the BCI's noisy, error-prone input recognition without undermining user agency.
Previous HCI research has studied how individuals with CP experienced aim assistance algorithms and networked games supporting different play styles via exercise bicycle inputs~\citep{hernandez_design_2014,hwang_how_2017}. However, they do not address the unique requirements of input with BCI technology, which provides no inherent feedback (e.g. a haptic, visual or auditory sensation~\citep{wensveen_interaction_2004}) when users attempt to control it. Understanding how individuals with CP experience game assistance in BCI contexts - where help is needed yet may compromise agency -  is therefore essential for advancing BCI training game designs and experiences.

To address this gap, we adopt a mixed-method protocol that triangulates field data collected in a clinic and a hospital with an exploratory proof-of-concept laboratory study. We gathered impressions from ten individuals with CP who attempted to control two BCI game prototypes with varying levels of sporadic in-game assistance. Combining structured questionnaires and semi-structured interviews allowed a preliminary quantitative analysis of preference for game assistance and examine key experiential dimensions like agency, frustration, motivation, and the perception of games for training. Our preliminary findings focus on how individuals with CP perceive game interactions and assistance, shedding light on the importance of tailoring training to accommodate individual preferences and experiences.

By evaluating the challenges and opportunities presented by BCI game training, we aim to provide insights valuable for HCI designers and BCI developers alike. Addressing these individual differences is essential for creating meaningful empowering experiences that can lead to significant improvements in training outcomes for individuals with CP.

\section{Background}

CP presents diverse motor and cognitive profiles that require training approaches tailored to individual functional abilities~\citep{reid_rehabilitation_2015}. Effective training/rehabilitation adheres to motor learning principles: repetitive training with high intensity~\citep{reid_rehabilitation_2015}, and exercises promoting activity-dependent plasticity~\citep{pascual-leone_modulation_1995} are particularly valuable. One such method is BCI training. BCIs translate brain activity into device commands and have consistently been shown to induce neural plasticity~\citep{niazi2022associative,mrachacz2016efficient} and functional improvements in stroke patients~\citep{biasiucci_brain-actuated_2018}. Similar functional changes are reported in individuals with CP~\citep{bobrov2020rehabilitation}. BCIs induce activity-dependent plasticity by pairing movement intentions from the motor cortex with sensory feedback from the affected limb, typically elicited via electrical stimulation, exoskeletons, or rehabilitation robots. The cortical activity and sensory feedback must coincide with very short latency~\citep{mrachacz2012precise} (less than 200 ms~\citep{grosse2011using}). This strict timing requires detecting movement intentions via electroencephalography (EEG). Individuals with CP can produce movement intentions that are detected from EEG with 70-80\% accuracy~\citep{jochumsen2025comparison, jadavji2023bci, jochumsen2018movement, daly2013control}.

Despite BCI performance being modest compared to keyboards or joysticks, BCIs offer value for individuals with severe impairments, both for training and as alternative input for games~\citep{maggiorini_evolution_2019}. However, in rehabilitation, poor BCI control can cause frustration~\citep{mane2020bci}, and the repetitive training can become monotonous, hence reducing training adherence. Games can reduce this frustration. Game assistance like fabricated input and performance accommodation techniques have been shown to increase agency and reduce frustration in stroke patients doing BCI training~\citep{jochumsen2022implementing, hougaard2022modulating}. Whether similar findings apply to people with CP remains an open question.

\subsection{Designing (Non-BCI) Games for CP adolescents}\label{sec:related-work}

Earlier work in HCI has explored how to build tailored games specifically for children with CP for other purposes than BCI training -- they investigated how to aid performance~\citep{hwang_how_2017}, support different play styles~\citep{hernandez_design_2014}, or create exergames for the home environment~\citep{albertsDesigningSmartphoneExergame2022}. 
Different aim assistance algorithms were explored in terms of how they narrowed the gap in player performance between children with and without CP~\citep{hwang_how_2017}. 
Using such aids can help overcome limitations due to motor impairments, providing more game design possibilities, for example developing games that incorporate high pace or time-sensitive actions~\citep{hernandez_designing_2013}. 
These games were developed for input devices such as exercise bicycles~\citep{hernandez_design_2014} or gamepads~\citep{hernandez_designing_2013}, depending on the child's motor and cognitive abilities.

\subsection{BCI Games for CP}\label{sec:rehab-games}
Only few studies have utilized BCI to control games for individuals with CP. To prove the feasibility of a BCI-based CP training paradigm,~\citeauthor{scherer_game-based_2015} proposed game-based BCI training to teach BCI control~\citep{scherer_game-based_2015}. They conducted two preliminary case studies showing CP users can achieve higher-than-chance cued BCI control and play Tic-Tac-Toe~\citep{scherer_game-based_2015,scherer_lets_2016}. \citeauthor{taherian_are_2017} studied six individuals with quadriplegic CP using a BCI system for puzzle games over three weeks~\citep{taherian_are_2017}. However, two participants discontinued after the first session due to BCI discomfort and lack of motivation. Recently, BCI games for CP have been developed in a BCI Game Jam and a tool for BCI game development~\citep{righttoplay2020, langflexible2020}. 
BCI game research for CP remains limited, though HCI collaboration could advance this area.

\subsection{Limitations to existing interdisciplinary work}\label{sec:gap}
Although biomedical engineering and HCI studies have each contributed important insights into BCI technology and game assistance for individuals with CP, an interdisciplinary usability evaluation of BCI games for this population is currently lacking. We address this gap with a user-centered investigation that specifically examines how individuals with CP experience assistance and agency in BCI games from three angles:

\begin{itemize}
\item \textbf{Technical angle (BCI):} How does characteristics of BCI (e.g. as unreliable input device) and characteristics of individuals with CP (e.g. atypical motor patterns) affect reliable training with BCI?\item \textbf{User experience:} How does agency and frustration influence people with CP when training with BCI games and what are the possible benefits to existing therapeutic practice? \item \textbf{Game assistance design:} How and when should BCI games provide assistance during BCI training without undermining users' sense of autonomy?
\end{itemize}

\section{Method}

\begin{figure}[h]
\centering
\includegraphics[width=0.9\textwidth]{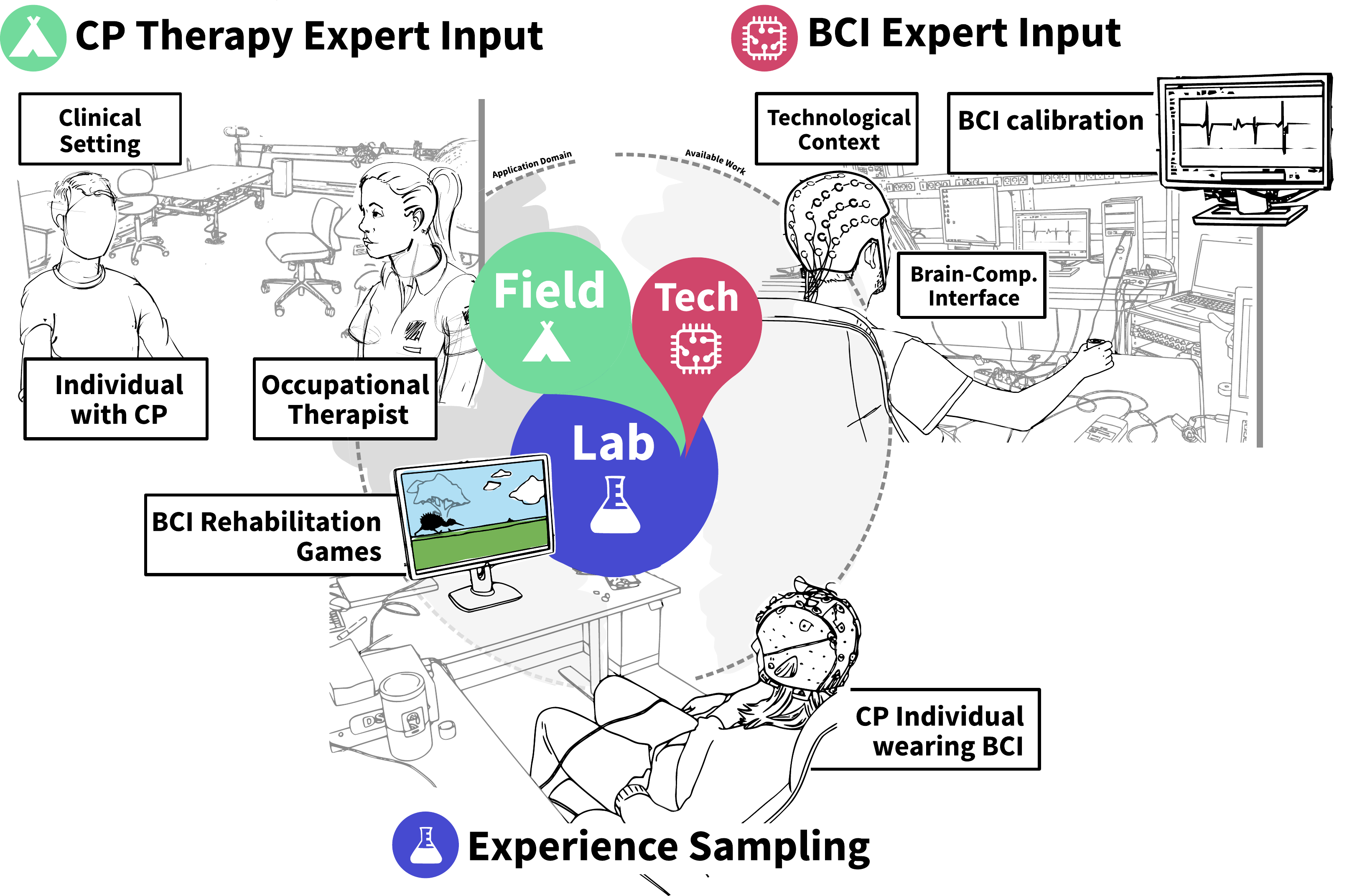}
\caption{Our mixed-method approach, visualized as a Development Oriented Triangulation framework~\citep{van_turnhout_design_2014} (center) with depictions of each activity. We gathered insights from real-world context through interviews in the clinical context, from conversations with biomedical engineering experts about brain-computer interfaces (BCIs), and from observing and interviewing individuals with cerebral palsy (CP) while they used BCIs in a lab setting}
\label{fig:bci-cp-mixed-method}
\end{figure}

We used a mixed-method approach depicted in Fig.~\ref{fig:bci-cp-mixed-method}, taking inspiration from \citeauthor{van_turnhout_design_2014}'s Development Oriented Triangulation (DOT) framework~\citep{van_turnhout_design_2014}. Our study synthesizes insights across field visits to the clinical therapy context, conversations with experts in the technological BCI context, and a laboratory study where individuals with CP got to experience BCI training first-hand.
This approach responded to constraints in our research niche: limited access to a sensitive clinical context and target group, and the nascent state of BCI technology outside the lab.
Section~\ref{sec:clincontext}--\ref{sec:labstudy} provides an overview of each type of activity and its goals.

\subsection{Visits of the Clinical Context}\label{sec:clincontext}
We conducted three clinical visits to gather qualitative insights on life with CP and training individuals with CP.

We obtained qualitative insights into life with CP and training individuals with CP through three clinical context visits. Objectives included 1) understanding the frame of CP therapy the healthcare system of Anonymized, 2) observe moment-to-moment therapeutic practice, and 3) uncover challenges and opportunities for game-based BCI training. Table~\ref{table:clinic-visits} summarizes each visit and the stakeholders present. Insights from these visits were documented by three note takers, and reported in Section~\ref{sec:opport} and Section~\ref{sec:design-cons}.

\begin{table}[!h]
\renewcommand{\arraystretch}{1.2}
\resizebox{\textwidth}{!}{
\begin{tabular}{p{2.8cm}p{9cm}p{3cm}}
Visit & Description & Stakeholders \\
\toprule
1. Observation of clinical therapy & We visited an occupational therapist in a clinic connected to a primary school in the municipality of Anonymized, to observe a 45-minute individual training session. & Occupational therapist, \newline adolescent with CP\\  
2. Observation in CP follow-up programs & We observed three follow-up meetings at a Danish hospital involving parents, the child, and CP-specialized staff (doctor and therapist). The meetings functioned as evaluations of the adolescent’s physical and mental health, abilities, and development (e.g. weight and height). & Doctor, \newline Occupational therapist, \newline Parents, \newline Adolescent with CP, \newline Child with CP \\  
3. Demonstrating BCI in clinical therapy & We revisited the clinic to demonstrate the concept of BCI therapy to the therapist, and invited her to provide thoughts on a BCI rehabilitation game concept through show-and-tell. & Occupational therapist \\
\toprule
\end{tabular}}
\caption{Overview of visits to the clinical context.}
\label{table:clinic-visits}
\end{table}

\subsection{Collaboration with BCI developers (biomedical engineers)}
Through a year-long collaboration between HCI researchers and biomedical engineering researchers, we obtained insight into BCI technology's use for rehabilitation purposes, specifically repetitive training based upon users attempting to perform hand movement. We were introduced to a wide range of headsets from single-electrode headsets like Myndband, to 16-channel headsets like OpenBCI and g.Nautilus.
The collaboration resulted in several game prototypes that explored ways to address key limitations of BCI as an input device, including:

\begin{enumerate}
\item The monotonous and repetitive nature of rehabilitation exercises can reduce engagement; BCI games can introduce variation and motivation.
\item Poor input recognition rates in BCI systems, e.g. from calibration, and the resulting reduction in agency can lead to frustration and into a vicious cycle of reducing recognition rates further; game design can help compensate for these limitations by maintaining users' sense of agency through meaningful feedback.

\end{enumerate}

The collaboration comprised monthly meetings  to coordinate research activities and address technical questions related to BCI interaction, supplemented by occasional sessions in a dedicated BCI laboratory for application demonstrations. Insights from this collaboration informed the identification of technical constraints and opportunities in Section~\ref{sec:design-cons} and ~\ref{sec:opport}.

\subsection{The Experience Sampling Laboratory Study}\label{sec:labstudy}

We designed a laboratory study that invited individuals with CP to try two different BCI-controlled game prototypes and share their experience. Our laboratory study's main aim was to shed light on the opportunities and the socio-technical challenges regarding gamified BCI training in individuals with CP.
Based on the technological context and clinical visits, the study took the first steps towards uncovering the experiential dimensions at play, when individuals with CP interact with BCI games. The study prioritized exploratory insight over statistical power, accepting limited control of variability due to constraints of the target group (availability, age, individual impairment) and the technology (variable BCI input recognition and game outcomes).

\subsubsection{Experimental Design}
The laboratory study followed a within-subject design with two BCI-controlled games being evaluated. Each participant tried the games multiple times, where each condition changed what type of game assistance was present, if any. To ensure that the game assistance was relevant to amplify the participant's experience, we limited
the BCI to provide participants with a maximum of 60\% input recognition, i.e. no more than 60\% of the attempted movements were used for the standard feedback of a successful attempt. We randomized the game order and condition order to minimize bias in our results from these variables. 
\subsubsection{Participants}
Ten participants with CP participated in the study (two female and eight male, mean age: 26±16 years), whose demographics can be seen in Table~\ref{table:study2-participants}. Each participant or their parents/caregivers gave their written consent before participating in the study. The study was approved by the North Denmark Region Committee on Health Research Ethics (N-20230025) and was conducted in accordance with the Declaration of Helsinki. 
Half of the participants had tried BCI in a previous study (P3, P4, P5, P8, P10) and one of the game prototypes (kiwi game).

\begin{table}[!h]
\renewcommand{\arraystretch}{1.1}
\resizebox{\textwidth}{!}{
\begin{tabular}{cllllp{5cm}p{5cm}}
PID & Gender & Age & Type & BCI Exp. & Functional Ability & Hand Ability \\
\toprule
6 & Male & 12 & Hemiplegia (right side) & - & No assistive devices. Most affected in right side, but used left during the game, due to high spasticity in right wrist. & Able to zip some jackets, but not all. Washes hair with unaffected hand. \\ 
7 & Male & 12 & Tetraparesis (right side) & - & Wheelchair, Affected in both sides. & Able to zip jacket. Washes hair with unaffected hand. \\ 
9 & Female & 12 & Hemiplegia (right side) & - & No assistive devices. A litlle spacticity on the right side & Able to zip some jackets, but not all.  Washes hair with both hands. \\  
2 & Female & 15 & Hemiplegia (right side) & - & No assistive devices, right arm lightly affected. & Able to zip jacket and tie shoes. Washes hair with unaffected hand with support from affected hand. \\  
1 & Male & 17 & Hemiplegia (right side) & - & No assistive devices, right arm is lightly affected. & Able to zip jacket, tie shoes and wash hair with both hands. \\  
5 & Male & 22 & Hemiplegia (right side) & Yes & No assistive devices Very well functioning, however experienced fatigue during the experiment. & Able to zip jacket, tie shoes and wash hair with both hands. \\ 
4 & Male & 32 & Hemiplegia (left side) & Yes & No assistive devices Left side quit affected - left wrist very spastic & Able to zip jacket. Washes hair with unaffected hand. \\
8 & Male & 34 & Tetraparesis (right side) & Yes & Wheelchair. Very affected in all 4 limbs, with high spacticity. & Afected in both hands, gets help to wash hair. \\ 
3 & Male & 51 & Diplegia (both legs) & Yes & Uses long stick as cane to assist walking. Left arm is lightly affected. Legs are much more affected compared to arms/hands. & Able to zip jacket, tie shoes and wash hair with both hands. \\ 
10 & Male & 56 & Hemiplegia (right side) & Yes & Light spasticity in right side, right wrist very fixed. & Washes hair with their unaffected hand. \\
\end{tabular}}
\caption{Participant demographics. The participants were asked the following questions to assess their abilities to do activities of daily living involving their hands and arms: 1) Can you zip your jacket? 2) can you tie your shoes? and 3) can you wash your hair?}
\label{table:study2-participants}
\end{table}

\subsubsection{Brain-Computer Interface}
The BCI used and 
OpenViBE's 
”Motor Imagery BCI” scenario~\citep{renard_openvibe_2010}, following previous work~\citep{jochumsen2022implementing}.
Continuous EEG was recorded using the g.NAUTILUSRESEARCH headset 16g.SCARABEO cap (g.tec medical engineering GmbH, Austria) from 16 channels (FP1, F3, FZ, F4, FC1, FCZ, FC2, C3, C1, CZ, C2, C4, CPZ, P3, PZ, and P4), grounded at AFZ and referenced to the right earlobe. The EEG was sampled at 250 Hz and transmitted via Bluetooth to a laptop. The EEG was bandpass filtered (8-30 Hz, 4th-order Butterworth) and a common spatially filter maximized 
the differences spectral power variance between movement and idle activity. Following this, a linear discriminant analysis classifier (LDA) was trained on the EEG calibration data using 5-fold cross-validation.

A participant-specific threshold (set by visual inspection) was validated in a brief (<5 min) online test session to balance true-positive and false-positive detections, with a bias towards reducing the number of true-positive and false-positive detections.

From a sliding window, the LDA continuously provided a probability output between 0 and 1 for each segment representing movement or idle activity at a rate of 16 Hz. 
For the participant to control the game, between three and seven consecutive outputs of the LDA were required to exceed the participant-specific threshold. By adjusting the number of consecutive output that needed to be above the threshold, the number of true-positives and false-positives could be adjusted accordingly.

\subsubsection{Experimental Setup}
\begin{figure}[h]
\centering
\includegraphics[width=\textwidth]{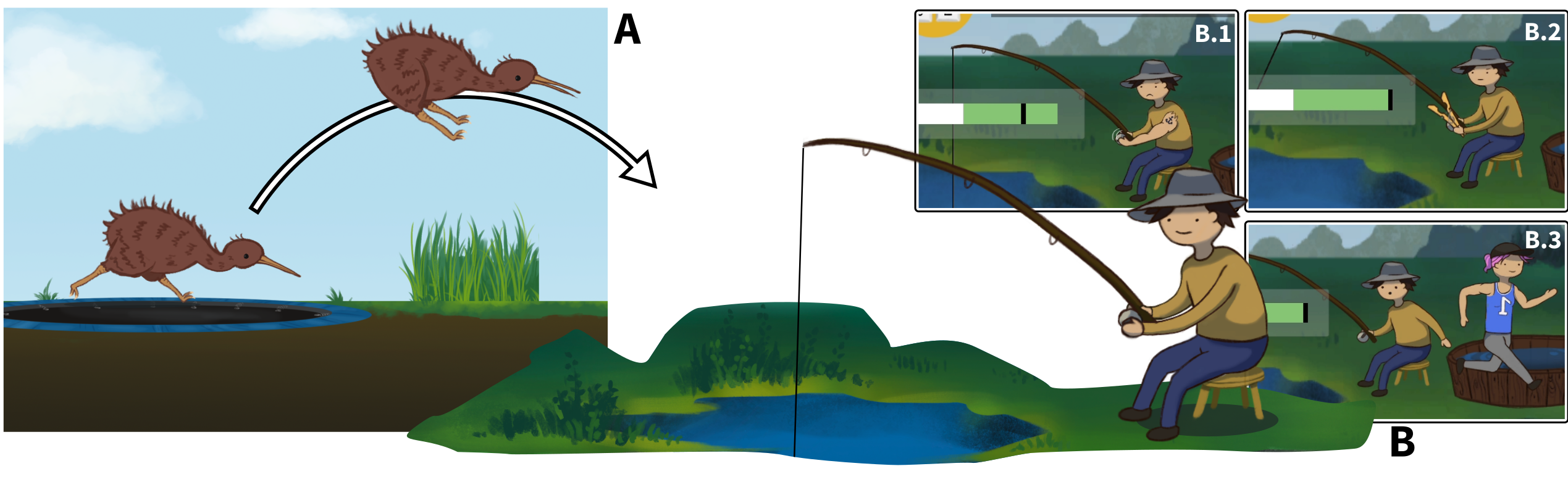}
\caption{In the kiwi game (A), users used BCI to help a kiwi jump over obstacles to rescue its babies in a race against an eagle, with occasional invisible assistance. In the fishing game (B), users controlled a fisherman to reel in fish with occasional visible assistance.}
\label{fig:bci-kiwi-help}
\end{figure}

The laboratory study employed two BCI games, shown in Fig.~\ref{fig:bci-kiwi-help}. In the kiwi game (A), users used the BCI to jump across obstacles, and were sometimes assisted through hidden help -- the game at random choose to fabricate successful input during 10-30\% of the attempts. In the fishing game (B), users controlled a fisherman to catch fish. B.1, B.2, and B.3 showcase three types of game assistance, the fisherman getting stronger (B.1), the fisherman preventing fish from escaping (B.2), and the fisherman receiving help from another person (B.3). In the kiwi game help was hidden from the player as fabricated inputs were not distinguishable from the users' successful attempts while in the fishing game all help was explicitly visible and explained. 

\subsubsection{Procedure}
First, the participant was seated and fitted with an EEG cap -- impedances and signal quality were then inspected. Calibration began, where the participants performed 30 wrist extensions with the hand of her/his choosing. The calibration was timed by a 4-second red-arrow cue to hold the movement (avoiding blinking or unintended muscle activity) followed by a 4-second "Rest" idle period. Each wrist extension was followed by an idle trial. After the calibration, participants were presented with the first game and asked to play it. After having played one session lasting about 2-3 minutes, the facilitator would show the participant a questionnaire and read each question one by one. When the participant provided a rating, the facilitator followed up by prompting them to elaborate their answer.

\subsubsection{Questionnaire}
We designed a questionnaire and semi-structured interview, to follow best practices for survey development for adolescents~\citep{Omrani_2019}. We extended and augmented a previously used survey~\citep{Hougaard_2021} and asked follow-up questions as a way to utilize qualitative interview methodology to have participants elaborate why and how they answered survey questions pertaining their perceived control, frustration and perception of receiving assistance from the game.  
To start a conversation about frustration, we asked participants to rate a single-item Likert scale asking \textit{''How frustrated did you feel with the game?''} based on previous literature~\citep{Miller_Mandryk_2016, Pedersen_2009,Lazar_2006,Zhou_2022}.
To measure agency, we adopted questions relating to agency from~\citet{Bergstrom_2022} such as \textit{''It felt like I was in control of the movements during the task''}, and \textit{''What is the degree of control you felt''}, available in Supplementary Material 1. In this paper we present preliminary findings from a thematic analysis of participants' verbal comments while answering these questions.

\subsubsection{Data Analysis}
\begin{figure}[!h]
\centering
\includegraphics[width=\textwidth]{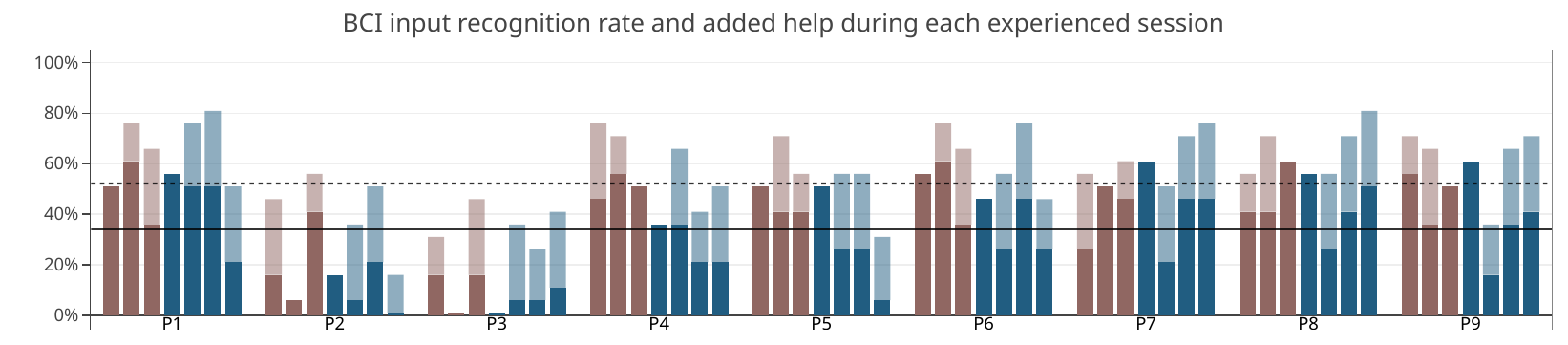}
\caption{
Between- and within-participant variation in BCI recognition rates and added game help. 
Each bar is a play-through of the kiwi game (brown) or fishing game (blue). Dark bars denote BCI recognition of attempted movements (solid line marks average). Light bars denote the added game help to boost training (dashed line marks average).}
\label{fig:recognition}
\end{figure}

We conducted open-coding thematic analysis of qualitative data, merging themes with field-observation themes where feasible.
Quantitative measures (BCI recognition rate and added help) were analyzed in R and reported as bar charts, means, and standard deviations.
Prior to synthesizing results, we analyzed individual recognition rates (the rate in which actions led to appropriate outcomes) which is summarized by the bar chart in Figure~\ref{fig:recognition}. Participants had 34\% recognition on average and received 18\% help in help conditions, providing on average 54\% non-failed outcomes. Recognition varied widely between and within participants (15\%–45\%); P2 and P3 had notably lower rates, which we considered when reviewing their responses.

\clearpage

\section{Results}
We synthesized results into three categories: 1)~Opportunities and concerns for BCI training (summarized by Table~\ref{table:opport-concerns-bci}), 2)~Design constraints for HCI (..Table~\ref{table:design-constraints}), and 3)~Expectations to help in training (..Table~\ref{table:help-control}).

\subsection{Opportunities and Concerns for BCI Training}\label{sec:opport}

\begin{table}[!h]
\renewcommand{\arraystretch}{1.6}
\resizebox{0.47\textwidth}{!}{
\begin{tabular}{p{0.5cm}p{7cm}}
\multicolumn{2}{l}{\Large \textbf{Opportunities for BCI Training}} \\
\toprule
\raisebox{-.75\height}{\includegraphics[width=0.50cm]{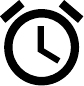}} & \textbf{Self-managed Home Training} \newline BCI games could in the future function as self-managed training assignments at home.  \\
\raisebox{-.75\height}{\includegraphics[width=0.40cm]{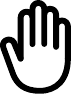}}  & \textbf{To Regain Limb Awareness} \newline Regaining control or sense of touch could act as motivations to use BCI-based training.  \\
\raisebox{-.75\height}{\includegraphics[width=0.60cm]{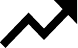}}  & \textbf{From Playful to Independence} \newline BCI games can play part in the transition from playful training to adult independence training.  \\
\raisebox{-.75\height}{\includegraphics[width=0.40cm]{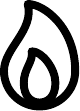}} & \textbf{Synergetic Training Warm-up} \newline BCI games could become functional warm-up to purposeful real-life activities with the therapist. 
\end{tabular}}
\hspace{0.5cm}
\resizebox{0.47\textwidth}{!}{
\begin{tabular}{p{0.50cm}p{7cm}}
\multicolumn{2}{l}{\Large \textbf{Potential Concerns For BCI Training}}\\
\toprule
\raisebox{-.75\height}{\includegraphics[width=0.50cm]{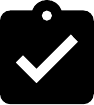}} & \textbf{Clarifying BCI Training Purpose} \newline BCI game training is on its own functional. It links only indirectly to practical outcomes.  \\
\raisebox{-.75\height}{\includegraphics[width=0.60cm]{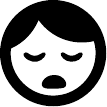}} & \textbf{Sustained engagement with training} \newline It's unclear how age-dependent fatigue affects motivation to train with BCI, particularly during adolescence. \\
\raisebox{-.75\height}{\includegraphics[width=0.40cm]{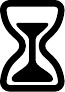}} & \textbf{Screen Time} \newline BCI rehabilitation games may require rebalancing users' screen time, particularly for adolescents. \\[4.8em]
\vphantom{1}
\end{tabular}}
\vspace{1em}
\caption{Overview of opportunities and concerns identified in brain-computer interface (BCI) training for people with cerebral palsy.}
\label{table:opport-concerns-bci}
\end{table}

\textbf{BCIs offer the opportunity to provide self-directed, at-home training}. 
Weekly training with a therapist is limited to few sessions per week depending on local policy -- for, example, in the Nordic municipality of Anonymized, this is limited to two sessions of 45 minutes. While between sessions, individuals are expected to engage in independent self-managed practice at home, adherence to this varies. 
Such training allows individuals to practice skills with occasional assistance from parents or at their own pace, fostering autonomy. During visit 1, the occupational therapist imagined that by leveraging engaging BCI games, individuals with CP can maintain their progress between formal training sessions -- for example, incorporating it as "homework" to create a continuous learning experience until the next supervised therapy session. 
The therapist imagined that \textbf{BCI training could become an opportunity for people who want to gain more awareness of their affected hand}. During visit 1 and 2, occupational therapists provided multiple nuances to what this means to individuals with CP, from wanting more muscle control, control over involuntary movements, to sense of touch, or feeling when others are touching them, depending on the person’s perceived deficits in their ability.
\textbf{BCI games create a middle ground in the transition from playful training to independence training.} At lower age, occupational therapists typically provide playful training (focusing on e.g. gameful and enjoyment training and establishing training habits), and transition to independence training during adolescence to focus on real-life skills such as zipping jackets, or regaining more limb control. Playful training engages younger children by making learning enjoyable, while independence training meets the practical needs of older adolescents. This shift supports skill development essential for autonomy, preparing adolescents for independent lives.
\textbf{The occupational therapist saw potential for utilizing BCI training as a warm-up activity in her own practice} before activity-based training. This could enable the therapist to multitask, extend sessions and free up more time for individualized training. 
\textbf{The occupational therapist's primary concern about utilizing BCI training in practice, was how to link it to practical outcomes}. 
Training should focus on meaningful exercises identified collaboratively with the adolescent, such as buttoning a jacket or cutting nails to enhance realism in practice. This contrasts with formulaic function-based training, which may lack clear goals beyond simply improving movement. While BCI training inherently is functional training, its effectiveness increases when integrated into activity-based contexts.
\textbf{Individuals with CP training for training's sake may develop cumulative fatigue}. Conversations between parents and therapists in our field visits indicated that exhaustion particularly can develop during adolescence.
To remedy these effects, occupational therapists suggested working with the parents to establish goal-based training that focuses on the adolescents' personal interests. \textbf{Parents voiced concerns about adolescents’ computer screen time}.
Adolescents with CP already spend considerable time on their computers when no other activities are scheduled.
Depending on how BCI game-based training is integrated into their daily routines, it may further increase overall screen time.

\subsection{Identified Design Constraints for HCI}\label{sec:design-cons}

\begin{table}[!h]
\renewcommand{\arraystretch}{1.6}
\resizebox{0.47\textwidth}{!}{
\begin{tabular}{p{0.75cm}p{6.75cm}}
\multicolumn{2}{l}{\Large \textbf{Designing for Individuals with Cerebral Palsy}} \\
\toprule
\raisebox{-.75\height}{\includegraphics[width=0.95cm]{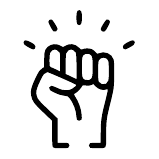}} & \textbf{Involuntary Movements} \newline Individuals with CP may tense muscles involuntarily, or grip but cannot let go. \\
\raisebox{-.75\height}{\includegraphics[width=0.95cm]{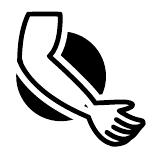}} & \textbf{Limb Awareness \& Mobility} \newline Affected limbs may not be sensitive to touch and may have reduced movement ability.  \\
\raisebox{-.75\height}{\includegraphics[width=0.95cm]{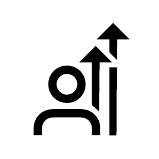}}  & \textbf{Help Trades Off Autonomy} \newline Occupational therapists rely on verbal hints and physical help, but wish to maintain people's autonomy during training. \\
\raisebox{-.75\height}{\includegraphics[width=0.95cm]{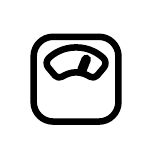}} & \textbf{Training Duration is Dynamically Negotiated} \newline Occupational therapist negotiates the duration of the training through verbal communication and by reading body language. \\
\end{tabular}}
\hspace{0.4cm}
\resizebox{0.47\textwidth}{!}{
\begin{tabular}{p{0.75cm}p{6.75cm}}
\multicolumn{2}{l}{\Large \textbf{Designing for Brain-Computer Interfaces}}\\
\toprule
\raisebox{-.75\height}{\includegraphics[width=0.95cm]{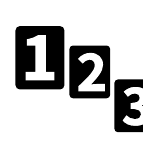}} & \textbf{BCI Training Requires Many Repetitions} \newline Effective BCI training for individuals with CP requires a large number of repetitions to induce plasticity (several thousands).  \\
\raisebox{-.75\height}{\includegraphics[width=0.95cm]{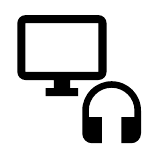}} & \textbf{BCI Feedback is Errornous} \newline Users experience difficulties in learning to control the BCI when presented with erroneous feedback system feedback on their performance. \\
\raisebox{-.75\height}{\includegraphics[width=0.95cm]{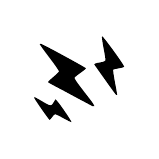}} & \textbf{BCIs are Susceptible to Noise} \newline Controlling a BCI demands concentration and the right mindset, with minimal external stimuli. \\[4.7em]
\vphantom{1}
\end{tabular}}
\caption{List of design constraints, synthesized from our field studies, that should be considered when designing interactive rehabilitation with brain-computer interfaces (BCIs) for individuals with cerebral palsy (CP).  }
\label{table:design-constraints}
\end{table}

The occupational therapist highlighted that \textbf{some individuals with CP experience involuntary movements} -- an arm may get in the way, and they might involuntarily contract a muscle without intentional control. Some find it easier to grip objects than to release them, because they struggle to let go. To avoid interference with gameplay, the occupational therapist suggested that game designers accommodate these challenges by handling input from adolescents with CP using strategies like specialized input hardware (e.g. switches instead of buttons) or error tolerance (e.g. enlarging input zones or extending time limits).
Some individuals with CP that could benefit from training with BCI,  \textbf{lack sensation and conscious awareness of their affected limbs}, making it difficult to control them (proprioception and kinesthetic awareness) or feel external contact (tactile discrimination). Designers may  need to pay attention to how they instruct tasks involving affected limbs, and identify ways to augment or virtualize tactile feedback.
The occupational therapist used verbal cues and physical assistance to guide adolescents through exercises, but also \textbf{wanted to preserve their autonomy during training}. Physical help, such as guiding hand movements can aid skill execution, but risks diminishing the trainee's autonomy. Observations revealed that therapists converse with the trainee during training to find the right autonomy balance, so the trainee becomes more independent over time. The therapist can then employ different levels of physical guidance depending on what is needed -- e.g. signifying where the hand should go (partial autonomy), or moving the hand on their behalf (no autonomy). The occupational therapist \textbf{negotiates the duration, structure, and agenda of each session} with the trainee to provide predictability. Observations from the field study showed that training duration was negotiated through verbal communication and by interpreting the trainee's body language, creating a collaborative decision-making when to end a given exercise. 
\textbf{BCI training requires many repetitions} to be effective and designers must therefore find ways to avoid decrease in motivation over time for individuals with CP.
Additionally, CP individuals' physical limitations or involuntary movements can hinder consistent execution of repetitive tasks, making it challenging to achieve the necessary number of repetitions for effective therapy which is expected to be in the range of thousands.
\textbf{BCI Feedback is errornous}, causing potential misalignment between attempted control of the BCI and the actual output. This impedes the learning for controlling the BCI. Developing a true sense of control and awareness over their actions can be challenging and hinder the overall training effectiveness.
\textbf{BCIs are susceptible to noise} disrupting accuracy of the decoding of EEG signals. This  leads to erroneous responses from the BCI system. The brain activity must be carefully recorded, sources of noise must be reduced as well as external stimuli and environmental interference to help users focus on controlling the BCI, which requires concentration and the right mindset.

\subsection{Expectations to Help and Impressions of BCI control}\label{sec:expectations-help}
\begin{figure}[!h]
\captionsetup{font=footnotesize}
\centering
\hspace{-12pt}
\hspace{-15pt}
\includegraphics[width=5.8cm]{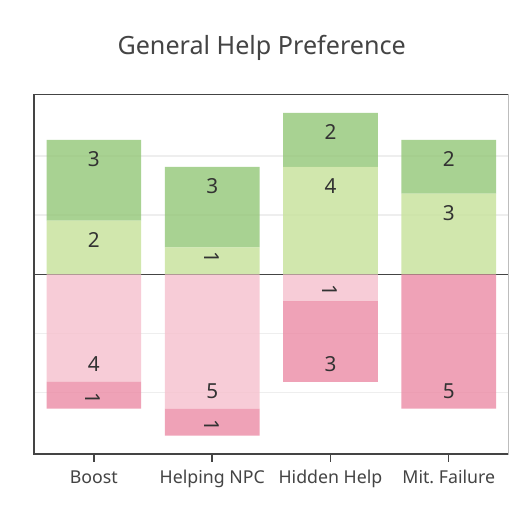}
\hspace{-13pt}
\includegraphics[width=5.8cm]{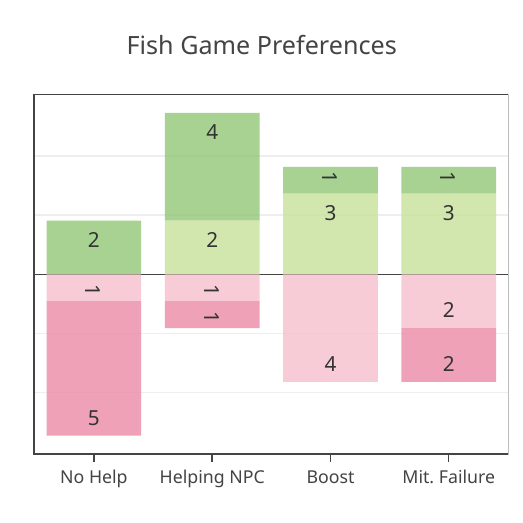}
\hspace{-13pt}
\vspace{3pt}
\includegraphics[width=4.95cm]{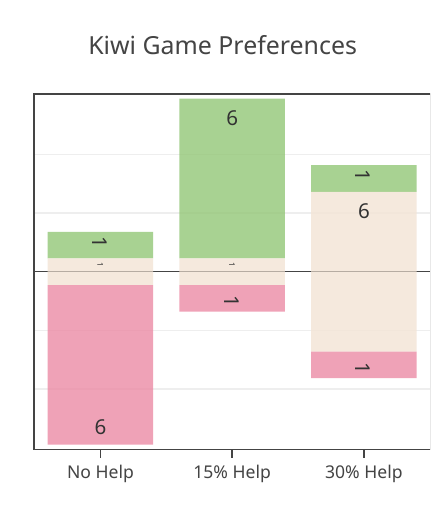}
\hspace{-13pt}
\caption{The participants ranked their help preference of the playthroughs of the kiwi (leftmost chart, from 1st to 3rd) and fishing game (middle chart, from 1st to 4th), and afterwards their general preference in help style (leftmost chart). Greener bars indicate higher preference, while redder bars indicate lower preference.)}
\label{fig:bar-ranking-preference}
\Description{....}
\end{figure}

Participants had mixed preferences for help as seen in Fig.~\ref{fig:bar-ranking-preference}. In the kiwi game, most preferred 15\% help, and fewest preferred full control without help.In the fishing game, four preferred the helping NPC (non-playable character) and five had least preference towards not receiving help. Overall, 3/10 participants preferred the Kiwi game, while 7/10 participants preferred the fish game.

Our laboratory study provided insight into how CP individuals viewed help in a live session, receiving it, and CP individuals impressions of controlling a BCI using attempted movements, summarized in Table~\ref{table:help-control}.

\begin{table}[!h]
\renewcommand{\arraystretch}{1.6}
\resizebox{0.47\textwidth}{!}{
\begin{tabular}{p{0.75cm}p{6.75cm}}
\multicolumn{2}{l}{\Large \textbf{Expectations To Help}} \\
\toprule
\raisebox{-.75\height}{\includegraphics[width=0.95cm]{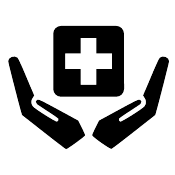}} & \textbf{Desire to Control Help} \newline Individuals with CP may want agency over the help they receive, e.g. to save help for the right moment. \\
\raisebox{-.75\height}{\includegraphics[width=0.95cm]{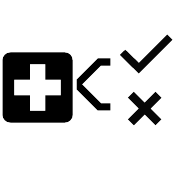}} & \textbf{Help Should Maintain Game Risk} \newline Help that progress the game while maintaining sense of risk were favored over help that only prevented negative consequences.  \\
\raisebox{-.75\height}{\includegraphics[width=0.95cm]{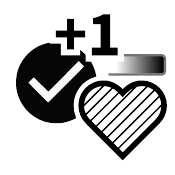}}  & \textbf{Visible Help Counters Monotony} \newline Visibly helping users increases variance in feedback outcomes, while invisible help can obfuscate poor outcomes. \\
\raisebox{-.75\height}{\includegraphics[width=0.95cm]{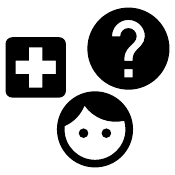}} & \textbf{Help Creates Doubt in Agency} \newline Visibily helping with actions reduced some users' autonomy. \\
\end{tabular}}
\hspace{0.4cm}
\resizebox{0.47\textwidth}{!}{
\begin{tabular}{p{0.75cm}p{6.75cm}}
\multicolumn{2}{l}{\Large \textbf{Impressions of BCI Control}}\\
\toprule
\raisebox{-.75\height}{\includegraphics[width=0.95cm]{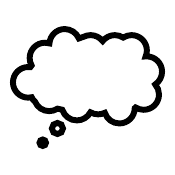}} & \textbf{Attributed Outcomes to Self-invented Actions} \newline Users self-invented ways to trigger BCI or suspected tampering with their control. \\
\raisebox{-.75\height}{\includegraphics[width=0.95cm]{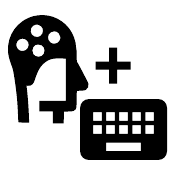}} & \textbf{Second Modality: Break or Distraction?} \newline Some users praised switching modalities because it gave training breaks. Others found switching modalities to distract
from the game experience. \\
\raisebox{-.75\height}{\includegraphics[width=0.95cm]{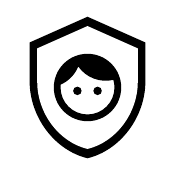}} & \textbf{Used to Frustration in Training} \newline Individuals with CP feel resilient to negative outcomes, and accustomed to train since childhood.  \\[6.1em]
\vphantom{1}
\end{tabular}}
\caption{Key expectations and impressions identified from our studies of how individuals with cerebral palsy (CP) view help and brain-computer interface (BCI) control.}
\label{table:help-control}
\end{table}

\subsubsection{Expectations to Help}
\textbf{Individuals with CP want to be in control of help}, such as using help at pivotal moments. Some participants felt more competent deciding when to seek help based on frustration, instead of receiving help at random from the game. As expressed by one participant, ``It should be possible to save the help for the right moment''. \textbf{Help should maintain game risk}. Participants preferred assistance that actively contributed to game progression while preserving a sense of risk, rather help that merely removed negative outcomes. One participant liked help that ``help you win and don't just prevent you losing something''. Another expressed that while help can be beneficial, ``it should also not be done too much, then it doesn't become fun to continue playing,'' reflecting the need for balance between support and challenge. Particularly P2, who had very poor recognition rates, did not appreciate help that only removed negative outcomes: ``The strong arm helped and the girl helped. The clothing peg did not help at all''.
\textbf{Visibly helping users introduces variation in gameplay}, which the participants preferred, when asked to choose between the fishing game and the more monotonous kiwi game. As one participant noted, “I don’t like experiencing 3-4 times in a row where [the kiwi] does not jump,” highlighting the frustration of repetitive outcomes. Conversely, another participant mentioned, “[The fishing game] was fun because [outcomes] changed each time,”. Invisible assistance in the kiwi game, allowed participants to perform without the pressure of recognizing their own deficiencies. As one participant noted, ``I think [the experience] is nicer if you don't know [that you are receiving help]''. Another participant expressed, ``I like that it is invisible, because you don't know when you receive it and you don't know that you are worse than you actually are''. At times, \textbf{invisible help created doubt in participants' control over outcomes}, when they felt it was unclear if actions were their own doing or caused by help in the kiwi game. Conversely, two participants preferred the kiwi game, because they felt the simplicity of outcomes (jumping/not jumping) made it more clear to them when the BCI had fallouts.

\subsubsection{Impressions of BCI Control}
Generally, the participant impressions of BCI control were mixed, one describing it as ``not a big brain twister'', and another as ``tiring for the head''. For some, the BCI response felt delayed with respect to the attempted movement, and for some they felt the repetitions fatigued their wrist.
\textbf{Users attributed outcomes to self‑invented triggers of the BCI}, such as a participant rationalizing that the system ``didn't react until the kiwi was halfway onto the trampoline'' and another blaming the cap for having "moments of dropout". One user suspected the study was tampering with their control to force the desired outcome -- ``you turned off control so it wouldn't jump''.
When participants were playing the fishing game, \textbf{the BCI was paired with a keyboard, asking the participants to switch between modalities.} One participants appreciated this alternation: “I actually felt it was fine to get a small break, so the wrist also had a little break.” However, the dual modality also introduced distraction -- a participant commented ``It was fun, but it also made it difficult to focus on the game itself. It became complicated to focus on the game due to both BCI and keyboard input.'' Individuals with CP reported that \textbf{frustration has become an expected part of training.} Initial annoyance gradually shifted to more resilient stance toward negative outcomes -- P3, who had particularly low recognition rates stated they were ``used to the kiwi not jumping,'' while another remarked, ``at first it was frustrating but then I got used to it.''

\section{Discussion}
Our findings provide an impression of how BCI training can co-exist with existing motor training therapy for individuals with CP. Our field visits uncovered potential pathways for BCI training to become a supportive training practice while our laboratory study demonstrated the feasibility of BCI-based game control with incorporated assistance. 

Regarding the \textbf{technical angle}, eight of ten participants achieved the expected level of BCI control over the game, though within-subject recognition varied. Two participants had lower recognition rates in all conditions, although the cause (BCI or atypical motor patterns) is unclear.

Regarding \textbf{user experience}, we learned that some individuals with CP themselves believe they develop resiliency to frustration over time, while therapists find that others are prone to training fatigue once reaching adolescence. So although BCI-based CP rehabilitation has functional resemblance to BCI-based in stroke rehabilitation, it's clear that the frustration and fatigue is different in nature. While interacting through BCI, some participants came up with other control methods or attributed control to other aspects of the interaction per instruction, suggesting that their extra effort increased their perceived system control. Such control illusions are not necessarily productive for BCI training, but are commonly found in skill situations, where apparent skill is an indicator of success likelihood~\citep{thompson_illusions_1998}.

Regarding \textbf{game assistance design}, the preference for maintaining a sense of risk when receiving help rather than eliminating negative consequences entirely, can seem counterintuitive to the very notion of assistance. It means that games need to consider how much assistance manipulates the game challenge and think about nuances around system-controlled assistance, such as visibility of assistance, and the current input recognition level. In~\citeauthor{hwang_how_2017}'s study of aim assistance, they recommended that help should be visible to the player, so that they can learn to adapt to the assistance and make use of it~\citep{hwang_how_2017}. Our results aligns with this recommendation, both from the perspective of the therapists' desire to maintain trainees' autonomy, and based on preference from participants, although in some cases invisible help was preferred by participants as a strategy to make their ability to control the games seem better (e.g. like training wheels).

\subsection{Limitations and Future Work}
The results presented are fundamentally limited by the number of involved individuals with CP (N=10) and experts, from the Nordic country of Anonymized. We primarily captured immediate post-session perceptions -- long-term adherence, motivation decay, and actual therapeutic transferability of the learned skills were not longitudinally tracked. 
Our study activities focused on controlled lab and clinic settings, which may hinder transferability of findings to BCI training in home environments.
Participants' impressions of help and BCI control in our laboratory study were only tested across the two BCI game prototypes and 4 types of in-game assistance they experienced. Although single-player and multi-player games previously promoted equally high training adherence~\citep{kaos_efficacy_2018}, the added social dimension of multi-player or co-operation play could be considered and compared to game-provided help as part of future studies.

We plan to collect additional data and conduct in-depth quantitative analyses of user agency and frustration, and their relationships with perceived difficulty, enjoyment, and boredom.  Specifically, we will examine whether agency can be attributed to discrete actions (e.g., reeling in a fish) versus broader tasks (e.g., catching a fish), and quantify the extent to which participants assign responsibility for negative outcomes to the input device (BCI), their own ability, or game events such as help.

\section{Conclusion}
In this article, we characterized the interaction constraints imposed by noisy BCI input and by the diverse motor, cognitive, and experiential needs of individuals with CP when undergoing game-based BCI motor training, which contribute to high variance in input recognition. To embrace such diversity, our methodology operated upon a mixed‑method protocol, synthesizing data from clinical visits with an exploratory proof‑of‑concept laboratory study.
When viewing BCI training in context of traditional therapy of CP, BCI training took a supplementary role and should be paired with an ulterior goal. We found that some individuals with CP believe they have a higher margin to frustration during training, which may present itself advantageous due to BCI's poor input recognition. Incorporating help may still alleviate monotony of BCI training, but we advice that designers should aim to maintain players' sense of risk and be aware of potential costs to players' sense of agency particularly when assistance is invisible.

\bibliographystyle{ACM-Reference-Format}
\bibliography{bibliography}
\end{document}